\documentclass[conference,10pt]{IEEEtran}
\IEEEoverridecommandlockouts
\usepackage{cite}
\usepackage{amsmath,amssymb,amsfonts}
\usepackage{algorithmic}
\usepackage{graphicx}
\usepackage{textcomp}
\usepackage{optidef}
\usepackage{xcolor}
\usepackage{overpic}

\def\imaginary{\mathsf{j}} 

\def\imaginary{\mathsf{j}} 
\def\Htran{\mbox{\tiny $\mathrm{H}$}}

\begin{document}

\title{Optimal Geometries of Dual-Polarized Arrays for Large Point-to-Point MIMO Channels}

\author{

 \IEEEauthorblockN{Amna Irshad, Emil Bj{\"o}rnson}
\IEEEauthorblockA{Department of Computer Science, KTH Royal Institute of Technology, Kista, Sweden \\
		{Email: amnai@kth.se, emilbjo@kth.se}
}
}

\maketitle

\begin{abstract}
Traditional point-to-point line-of-sight channels have rank $1$, irrespective of the number of antennas and array geometries, due to far-field propagation conditions. By contrast, recent papers in the holographic multiple-input multiple-output (MIMO) literature characterize the maximum channel rank that can be achieved between two continuous array apertures, which is much larger than $1$ under near-field propagation conditions. In this paper, we maximize the channel capacity between two dual-polarized uniform rectangular arrays (URAs) with discrete antenna elements for a given propagation distance. In particular, we derive the antenna spacings that lead to an ideal MIMO channel where all singular values are as similar as possible. We utilize this analytic result to find the two array geometries that respectively minimize the aperture area and the aperture length.
\end{abstract}
\begin{IEEEkeywords}
Near-field, MIMO system, channel capacity.
\end{IEEEkeywords}

\section{Introduction}
The capacity requirements on wireless communication links continue to grow, and we cannot cater to them by indefinitely increasing the spectral bandwidth, which is a limited resource. Alternatively, multiple-input multiple-output (MIMO) technology can be used to improve capacity. MIMO enables spatial multiplexing in rich multi-path propagation scenarios \cite{Telatar1999a}, where the capacity grows proportionally to the minimum of the number of transmit and receive antennas (i.e., the rank of the MIMO channel matrix). However, sixth-generation (6G) mobile systems operating at mmWave and sub-terahertz frequencies will feature line-of-sight (LOS) dominant channel conditions \cite{Rappaport2019}. Free-space point-to-point LOS MIMO channels were traditionally viewed to have rank $1$ \cite[Sec.~7.2.3]{Tse2005a} since only a single planar wave can be transferred from the transmitter to the receiver.
However, a higher rank can also be achieved in LOS MIMO by exploiting spherical wavefronts when operating in the radiative near-field region \cite{Decarli2021a}, which can have an extensive range in 6G bands since it is inversely proportional to the wavelength \cite{Ramezani2023a}.

The maximum rank that a continuous \emph{holographic} array can achieve was characterized in \cite{Hu2018a,Pizzo2020a}, but the maximum is not achieved by LOS MIMO links.
Moreover, it is not only the rank that determines the channel capacity but also the condition number of the channel matrix, which should ideally be one.
It was shown in \cite{multi2011,DoHeedong2020ColM} how the antenna spacing in two uniform linear arrays (ULAs) can be optimized to achieve such an ideal LOS MIMO channel.
The case of uniform rectangular arrays (URAs) was considered in \cite{LarssonP.2005,Zhou2012AADf}. However, these prior works are limited to single-polarized arrays, although practical systems generally utilize dual polarization.

In this paper, we consider a LOS MIMO channel between two dual-polarized URAs, with imperfect isolation \cite{Nabar2002a,coldrey2008modeling,Emil2017}. We analytically derive the horizontal and vertical antenna spacing that maximizes the MIMO channel capacity in the spatial multiplexing regime where the signal-to-noise ratio (SNR) is large. We provide a closed-form expression for the maximum attainable capacity. The results are qualitatively and quantitatively different than in the prior work due to imperfect polarization isolation \cite{LarssonP.2005,Zhou2012AADf}. We utilize the new analytic results to optimize the array geometries further to minimize either the area of the URA or the maximum aperture length. The analytical results are corroborated numerically.

\section{System Model}

\begin{figure}[t!]
			\begin{overpic}[width=\columnwidth,tics=10]{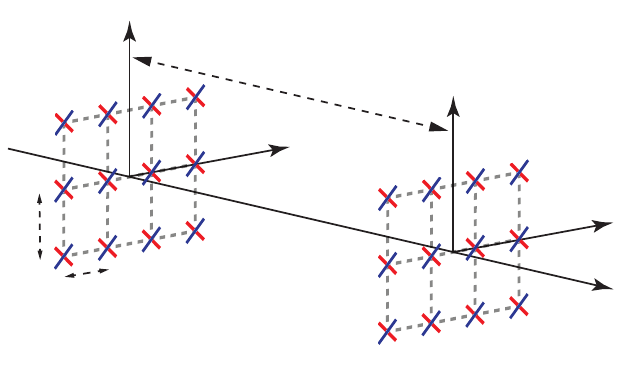}
                \put(11,10.5){\small $\Delta_{\rm h}$}
                \put(1,45.5){\footnotesize $(i(m),j(m))$}
				\put(0.5,21){\small $\Delta_{\rm v}$}
    		    \put(45,43.5){\small $d$}
    		    \put(10,5){\small Transmitter}
    		    \put(76,2){\small Receiver}
				\put(51,25){\footnotesize $(1,1)$}
				\put(51,14){\footnotesize $(1,2)$}
				\put(51,4){\footnotesize $(1,3)$}
				\put(62.5,30.5){\footnotesize $(2,1)$}
                \put(72,38){\footnotesize $(i(k),j(k))$}
				\put(71,32){\footnotesize $(3,1)$}
				\put(79,34){\footnotesize $(4,1)$}
			\end{overpic}   
\caption{A LOS channel between two dual-polarized arrays separated by a distance $d$. In this example, $M_{\rm h}=4$ and $M_{\rm v}=3$, while the horizontal/vertical indices are shown at the receiver side.}
\label{fig}
\end{figure}
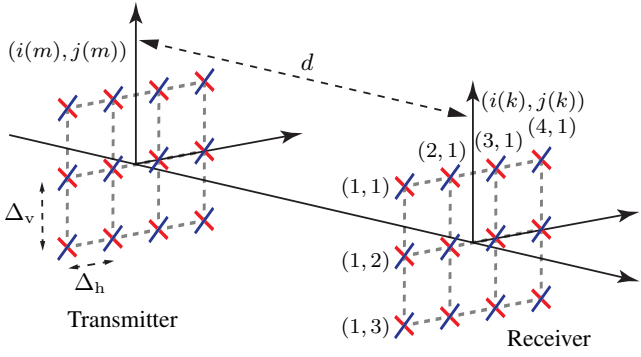

We consider a point-to-point free-space LOS channel between two URAs. The arrays have aligned broadside directions and are separated by a distance $d$, as illustrated in Fig.~\ref{fig}. The URAs are identically arranged with $M_{\rm v}$ vertically stacked rows and $M_{\rm h}$ antennas per horizontal row, which makes the total number of antenna locations $M=M_{\rm v}M_{\rm h}$. The vertical and horizontal antenna spacings are denoted by $\Delta_{\rm v}$ and $\Delta_{\rm h}$, respectively. Each antenna is dual-polarized, thus, it consists of two co-located elements with orthogonal polarization dimensions (e.g., slanted $\pm 45^\circ$) \cite{Emil2017}. Hence, the total number of antenna elements per array is $2M$.

The antenna locations in each array are numbered row by row from $1$ to $M$, but when computing the propagation distances, it is convenient to extract an antenna's horizontal and vertical indices. For a given antenna index $m \in \{ 1,\ldots,M\}$, the 
horizontal index can be calculated as \cite[Sec.~7.3]{Emil2017}
\begin{align} \label{eq:horizontal-index}
  i(m) = m- M_{\rm h} \left\lfloor\frac{m-1}{M_{\rm h}}\right\rfloor \in \{1,\ldots,M_{\rm h}\},
  \end{align}
  where $\lfloor \cdot \rfloor$ truncates the arguments to the closest smaller integer.
 The vertical index is similarly computed as
  \begin{align} \label{eq:vertical-index}
 j(m) =  1+ \left\lfloor\frac{m-1}{M_{\rm h}}\right\rfloor
  \in \{1,\ldots,M_{\rm v}\}.
 \end{align}
Using this notation, the distance between transmit antenna location $m$ and receive antenna location $k$ is obtained as
\begin{equation}\label{eq:3}
d_{m,k}=\sqrt{d^2+\big( i(m)-i(k) \big)^2 \Delta_{\rm h}^2 +\big( j(m)-j(k) \big)^2 \Delta_{\rm v}^2}.
 \end{equation}
The value depends on the horizontal/vertical antenna spacings.

\subsection{Channel matrix model for dual-polarized URAs}

In a LOS scenario with single-polarized antennas, the MIMO channel matrix $\mathbf{H}_{\rm u} \in \mathbb{C}^{M \times M}$ can be expressed as
\begin{equation} \label{eq:Hu}
\mathbf{H}_{\rm u}=    
\begin{bmatrix}
   \sqrt{\beta_{1,1}} e^{-\imaginary2\pi\frac{d_{1,1}-d}{\lambda}} & \cdots&\sqrt{\beta_{1,M}}e^{-\imaginary2\pi\frac{d_{1,M}-d}{\lambda}}\\
  \vdots &\ddots &\vdots \\
  \sqrt{\beta_{M,1}}e^{-\imaginary2\pi\frac{d_{M,1}-d}{\lambda}} & \cdots&\sqrt{\beta_{M,M}}e^{-\imaginary2\pi\frac{d_{M,M}-d}{\lambda}}\\
\end{bmatrix},
\end{equation}
where $d$ is the reference distance for the phase shifts, $\lambda$ denotes the wavelength, and the channel gain between isotropic antennas at transmitter  location $m$ and receiver location $k$ is
\begin{equation}
\beta_{m,k} = \left( \frac{\lambda}{4 \pi d_{m,k}} \right)^2.
\end{equation}
However, we consider two dual-polarized antenna arrays. In the ideal case when the orthogonal polarizations are perfectly isolated, the corresponding channel matrix $\mathbf{H}_{\rm d} \in \mathbb{C}^{2M \times 2M}$ can be expressed as
\begin{equation}\label{eq:H_d_def}
    \mathbf{H}_{\rm d}=\begin{bmatrix}
        \mathbf{H}_{\rm u} & \mathbf{0}\\
        \mathbf{0} & \mathbf{H}_{\rm u}
    \end{bmatrix}
    = \mathbf{I}_2 \otimes \mathbf{H}_{\rm u},
\end{equation}
where $\mathbf{I}_2$ is the $2 \times 2$ identity matrix and $\otimes$ denotes the Kronecker product. We achieve this matrix formulation by letting rows/columns $1,\ldots,M$ consider all elements having the first polarization and rows/columns $M+1,\ldots,2M$ having the second polarization. The $M \times M$ matrices with zeros in \eqref{eq:H_d_def} represent perfect isolation between the polarizations.

Although the polarizations of signals are maintained in free-space propagation, cross-talk generally appears in the transceiver hardware \cite{Nabar2002a,coldrey2008modeling,Emil2017}. This is referred to as imperfect cross-polar discrimination (XPD). We assume that each transmit antenna element radiates a fraction $(1-\gamma)$ of its power into the intended polarization and the remaining fraction $\gamma$ into the opposite polarization. The parameter $\gamma \in [0,1]$ specifies the impurity of the antenna, where $\gamma=0$ in the ideal case.

Motivated by hardware symmetry, we further consider that each receive antenna element captures a fraction $(1-\gamma)$ of the incident power of the signal  having the intended polarization and a fraction $\gamma$ of the power from the opposite polarization. Consequently, when considering a pair of dual-polarized transmit and receive antennas, 
a fraction
\begin{equation} \label{eq:polarization1}
(1-\gamma)^2 + \gamma^2  = 1 - 2 (1-\gamma) \gamma 
\end{equation}
of the signal power reaches the receiver with the correct polarization. Moreover, a fraction
\begin{equation} \label{eq:polarization2}
(1-\gamma) \gamma +  \gamma (1-\gamma)= 2 (1-\gamma) \gamma
\end{equation}
leaks into the opposite polarization, either at the transmitter or the receiver.
Note that the sum of \eqref{eq:polarization1} and \eqref{eq:polarization2} is $1$, thus, the total signal power is maintained irrespective of the value of $\gamma$.

By introducing the notation $\kappa=2 (1-\gamma) \gamma$, we generalize the dual-polarized channel matrix in \eqref{eq:H_d_def} as 
\begin{align} \nonumber
    \mathbf{H}_{\rm d}&=\begin{bmatrix}
        \sqrt{1-\kappa}\mathbf{H}_{\rm u} & \sqrt{\kappa} \mathbf{H}_{\rm u}\\
        \sqrt{\kappa} \mathbf{H}_{\rm u} & \sqrt{1-\kappa}\mathbf{H}_{\rm u}
    \end{bmatrix} \\ \label{eq:H_d_def_kappa}
    &= \underbrace{\begin{bmatrix}
        \sqrt{1-\kappa} & \sqrt{\kappa}\\
        \sqrt{\kappa} & \sqrt{1-\kappa}
    \end{bmatrix}}_{=\mathbf{K}} \otimes \mathbf{H}_{\rm u}.
\end{align}
This is the dual-polarized channel model that we will consider in the remainder of this paper.
Since the Frobenius norm of a Kronecker product is the product of the Frobenius norms, it follows that the channel matrix has the norm
\begin{equation}
\| \mathbf{H}_{\rm d} \|_{\rm F}^2 = \| \mathbf{K} \|_{\rm F}^2  \| \mathbf{H}_{\rm u} \|_{\rm F}^2 = 2 \sum_{m=1}^{M} \sum_{k=1}^{M} \beta_{m,k}.
\end{equation}
This value is independent of the XPD in $\mathbf{K}$ and the phase shifts of the individual elements in $\mathbf{H}_{\rm u}$. Nevertheless, the MIMO channel capacity depends on these parameters because they determine how the value of $\| \mathbf{H}_{\rm d} \|_{\rm F}^2$ is divided between the eigenvalues of $\mathbf{H}_{\rm d}^{\Htran}\mathbf{H}_{\rm d}$. The objective of the paper is to identify the planar array geometry (e.g., the antenna spacing) that maximizes the channel capacity in high-SNR scenarios.

\section{Optimal Antenna Spacing at High SNR}

The MIMO channel capacity depends on the eigenvalues of $\mathbf{H}_{\rm d}^{\Htran}\mathbf{H}_{\rm d}$ \cite{Telatar1999a} and the sum of the eigenvalues is $\| \mathbf{H}_{\rm d} \|_{\rm F}^2$. The high-SNR capacity is maximized when the eigenvalues are as equal as possible.
To analytically derive the antenna spacing that maximizes capacity, we need to simplify the channel matrix expression in \eqref{eq:H_d_def_kappa}. 
The diagonal in each array has the length
\begin{equation}
    D = \sqrt{(M_{\rm h}-1)^2 \Delta_{\rm h}^2 +(M_{\rm v}-1)^2 \Delta_{\rm v}^2}.
\end{equation}
In typical propagation scenarios for which $d \geq 2D$ \cite{Ramezani2023a}, the channel gain is nearly the same between all antenna locations:
\begin{equation}
    \beta_{m,k} \approx \beta = \left( \frac{\lambda}{4 \pi d} \right)^2. \label{eq:first-approx}
\end{equation}
Furthermore, the Taylor approximation $\sqrt{1+x^2/d^2}\approx 1+\frac{x^2}{2d^2}$ is tight for $x \leq D$ when $d \geq 2D$. Consequently, the following approximation is also tight:
\begin{align} \nonumber 
d_{m,k} &= d \sqrt{1+\frac{\big( i(m)-i(k) \big)^2 \Delta_{\rm h}^2 +\big( j(m)-j(k) \big)^2 \Delta_{\rm v}^2}{d^2}} \\
&\approx d+\frac{ \delta_{m,k} }{2d}, \label{eq:second-approx}
 \end{align}
where $\delta_{m,k} = (i(m)-i(k))^2 \Delta_{\rm h}^2 +( j(m)-j(k) )^2 \Delta_{\rm v}^2$.
By utilizing \eqref{eq:first-approx} and \eqref{eq:second-approx}, we can tightly approximate
$\mathbf{H}_{\rm u}$ in \eqref{eq:Hu} as
\begin{equation} \label{eq:Hu_approx}
\mathbf{H}_{\rm u} \approx  \tilde{\mathbf{H}}_{\rm u} =  \sqrt{\beta}
\begin{bmatrix}
    e^{-\imaginary\pi\frac{\delta_{1,1}}{d\lambda}} & \cdots& e^{-\imaginary\pi\frac{\delta_{1,M}}{d\lambda}}\\
  \vdots &\ddots &\vdots \\
   e^{-\imaginary\pi\frac{\delta_{M,1}}{d\lambda}} & \cdots& e^{-\imaginary\pi\frac{\delta_{M,M}}{d\lambda}}\\
\end{bmatrix}.
\end{equation}
Based on this expression, it follows that $\| \tilde{\mathbf{H}}_{\rm u} \|_{\rm F}^2 = \beta M^2$ is the sum of the eigenvalues of $\tilde{\mathbf{H}}_{\rm u}^{\Htran}\tilde{\mathbf{H}}_{\rm u}$.
 
\subsection{Preliminaries for ULAs}
The optimal antenna spacing with horizontal single-polarized ULA (i.e., $M_{\rm v}=1$) was characterized in \cite{multi2011,DoHeedong2020ColM}. We will briefly summarize this result because it will  later be reused.
If we can find an antenna spacing $\Delta_{\rm h}$ so that
 $\tilde{\mathbf{H}}_{\rm u}^{\Htran}\tilde{\mathbf{H}}_{\rm u} = \beta \mathbf{I}_M$, then all eigenvalues are equal and the capacity is maximized. By utilizing \eqref{eq:Hu_approx} but dropping the horizontal subscript so that $\delta_{m,k} = (m-k)^2 \Delta^2$, we obtain that the $(l,k)$th entry is
   \begin{equation}\label{eq:6}
   [\tilde{\mathbf{H}}_{\rm u}^{\Htran}\tilde{\mathbf{H}}_{\rm u}]_{l,k} =
   \beta \sum_{m=1}^{M} e^{\imaginary\frac{\pi}{d\lambda}(\delta_{m,l}-\delta_{m,k})},
   \end{equation}
which is $\beta M$ if $l=k$. The  magnitude of an off-diagonal entry can be simplified using the geometric series formula as
\begin{align} \nonumber
   \beta \left|\sum_{m=1}^{M} e^{\imaginary\frac{\pi}{d\lambda}(\delta_{m,l}-\delta_{m,k})} \right| &= \beta \left|\sum_{m=1}^{M} e^{\imaginary\frac{2 \pi (l-k) \Delta^2}{d\lambda} } \right|
   \\ &
   = \beta \left| \frac{1-e^{\imaginary\pi\frac{2M(l-k)\Delta^2}{\lambda d}}}{1-e^{\imaginary\pi\frac{2(l-k)\Delta^2}{\lambda d}}} \right|, \label{eq:ULA-geometric-series}
   \end{align}
which is zero if $\frac{M\Delta^2}{\lambda d}=1$ and $M>1$. By solving for $\Delta$, we obtain the  optimal antenna spacing as 
\begin{equation}\label{eq:7}
    \Delta=\sqrt{\frac{\lambda d}{M}}.
\end{equation}

\subsection{Optimal spacing for dual-polarized URAs}

We will now derive the optimal horizontal and vertical antenna spacings for the dual-polarized MIMO channel. 
Based on the tight approximation in \eqref{eq:Hu_approx}, we can express the dual-polarized channel matrix in \eqref{eq:H_d_def_kappa} as
$\tilde{\mathbf{H}}_{\rm d} = \mathbf{K} \otimes \tilde{\mathbf{H}}_{\rm u}$.
By utilizing properties of the Kronecker product, we obtain
\begin{equation}\label{Hdual}
    \tilde{\mathbf{H}}_{\rm d}^{\Htran}\tilde{\mathbf{H}}_{\rm d}= (\mathbf{K} \otimes \tilde{\mathbf{H}}_{\rm u})^{\Htran}(\mathbf{K} \otimes \tilde{\mathbf{H}}_{\rm u})=(\mathbf{K}^{\Htran}\mathbf{K}) \otimes (\tilde{\mathbf{H}}_{\rm u}^{\Htran}\tilde{\mathbf{H}}_{\rm u}).
\end{equation}
The eigenvalues of this matrix are obtained as the pairwise products of the eigenvalues of $\mathbf{K}^{\Htran}\mathbf{K}$ and $\tilde{\mathbf{H}}_{\rm u}^{\Htran}\tilde{\mathbf{H}}_{\rm u}$.
The first of these has the eigenvalue decomposition
\begin{align}\nonumber
    \mathbf{K}^{\Htran}\mathbf{K} &= \begin{bmatrix}
        1 & 2\sqrt{\kappa}\sqrt{1-\kappa}\\
        2\sqrt{\kappa}\sqrt{1-\kappa} & 1
    \end{bmatrix} \\ \label{eq:15} &=
    \frac{1}{2} \begin{bmatrix}
        1 & -1\\
        1 & 1
    \end{bmatrix}
    \begin{bmatrix}
        \mu_1 & 0\\
        0 & \mu_2
    \end{bmatrix}
\begin{bmatrix}
        1 & 1\\
        -1 & 1
    \end{bmatrix},
\end{align}
where the eigenvalues are
\begin{align} \label{eq:eigenvalues1}
\mu_1 &= 1+2\sqrt{(1-\kappa)\kappa}, \\
\mu_2 &= 1-2\sqrt{(1-\kappa)\kappa}. \label{eq:eigenvalues2}
\end{align}
These eigenvalues depend on the XPD parameter $\kappa$, but not on the antenna spacing so we cannot optimize them. Hence, we must focus on making all the eigenvalues of $\tilde{\mathbf{H}}_{\rm u}^{\Htran}\tilde{\mathbf{H}}_{\rm u}$ being equal, which happens when it is a scaled identity matrix. The $(l,k)$th entry of the matrix is
\begin{align} \nonumber
&[\tilde{\mathbf{H}}_{\rm u}^{\Htran}\tilde{\mathbf{H}}_{\rm u}]_{l,k} 
= \beta \sum_{m=1}^{M} e^{\frac{\imaginary \pi }{d \lambda}(\delta_{m,l}-\delta_{m,k}) }
 \\ \nonumber
    &= A_{l,k} \sum_{m=1}^{M} e^{\frac{\imaginary 2\pi }{d \lambda}i(m)[i(k)-i(l)]\Delta_{\rm h}^{2}}  e^{\frac{\imaginary 2\pi }{d \lambda}j(m)[j(k)-j(l)]\Delta_{\rm v}^{2}}
 \\ &
    =A_{l,k} \sum_{m_{\rm h}=1}^{M_{\rm h}} e^{\frac{\imaginary 2\pi }{d \lambda}m_{\rm h}[i(k)-i(l)]\Delta_{\rm h}^{2}} \sum_{m_{\rm v}=1}^{M_{\rm v}} e^{\frac{\imaginary 2\pi }{d \lambda}m_{\rm v}[j(k)-j(l)]\Delta_{\rm v}^{2}}, \label{eq:offdiagonals_URA_Hu}
\end{align}
where the following scalar is independent of $m$:
\begin{align}
    A_{l,k}= \beta e^{\frac{\imaginary \pi}{d \lambda}[(i(l)^2-i(k)^2)\Delta_{\rm h}^2+(j(l)^2-j(k)^2)\Delta_{\rm v}^2] }.
\end{align}
The last expression in \eqref{eq:offdiagonals_URA_Hu} is obtained by summing over the horizontal and vertical dimensions separately.
The diagonal entries equals $\beta M$, while the off-diagonal entries (i.e., $l \neq k$) have the magnitude
    \begin{align} \nonumber
&\beta \left| \sum_{m_{\rm h}=1}^{M_{\rm h}} e^{\frac{\imaginary 2\pi }{d \lambda}m_{\rm h}[i(k)-i(l)]\Delta_{\rm h}^{2}} \right| \left| \sum_{m_{\rm v}=1}^{M_{\rm v}} e^{\frac{\imaginary 2\pi }{d \lambda}m_{\rm v}[j(k)-j(l)]\Delta_{\rm v}^{2}} \right| \\
&=    
    \beta \left| \frac{1-e^{\imaginary\pi\frac{2M_{\rm h}(i(l)-i(k))\Delta_{\rm h}^2}{\lambda d}}}{1-e^{\imaginary\pi\frac{2(i(l)-i(k))\Delta_{\rm h}^2}{\lambda d}}} \right| \left| \frac{1-e^{\imaginary\pi\frac{2M_{\rm v}(j(l)-j(k))\Delta_{\rm v}^2}{\lambda d}}}{1-e^{\imaginary\pi\frac{2(j(l)-j(k))\Delta_{\rm v}^2}{\lambda d}}} \right|,
\end{align}
where the equality follows from using the classical geometric series formula. Each of these terms has the same structure as in \eqref{eq:ULA-geometric-series}, but only depends on either the horizontal or vertical spacing. Hence, the optimal antenna spacings must satisfy $\frac{M_{\rm h}\Delta_{\rm h}^2}{\lambda d}=1$  and $\frac{M_{\rm v}\Delta_{\rm v}^2}{\lambda d}=1$. By solving for $\Delta_{\rm h}$ and $\Delta_{\rm v}$, we obtain the solutions
\begin{equation}\nonumber
    \Delta_{\rm h}=\sqrt{\frac{\lambda d}{M_{\rm h}}},
     \quad \Delta_{\rm v}=\sqrt{\frac{\lambda d}{M_{\rm v}}}, 
\end{equation}
where the optimal spacing in one dimension only depends on the number of antennas in that dimension (and the propagation distance $d$ and wavelength $\lambda$).
In the special case of  $M_{\rm h} = M_{\rm v}$, we consequently get $\Delta_{\rm h} = \Delta_{\rm v}$. Hence, if there is the same number of antennas per dimension, a uniform square array (USA) with $\Delta_{\rm h}=\Delta_{\rm v}=\sqrt{\frac{\lambda d}{M_{\rm h}}}$ is optimal.

Interestingly, the optimal spacing is the same with our dual-polarized array as for the single-polarized arrays considered in \cite{LarssonP.2005,Zhou2012AADf}, irrespective of the XPD. However, the resulting eigenvalues of the channel matrix are different and depend on $\kappa$.
Since $\tilde{\mathbf{H}}_{\rm u}^{\Htran}\tilde{\mathbf{H}}_{\rm u}=\beta M \mathbf{I}_M$ with the optimal spacing, it follows that $\tilde{\mathbf{H}}_{\rm d}^{\Htran}\tilde{\mathbf{H}}_{\rm d} = \mathbf{K}^{\Htran}\mathbf{K} \otimes \beta M \mathbf{I}_M$ has $M$ eigenvalues that equal $\mu_1\beta M$ and $M$ eigenvalues that equal $\mu_2 \beta M$. 

\subsection{Capacity with optimized dual-polarized planar arrays}

For a given maximum transmit power $P$ and noise variance $\sigma^2$, the channel capacity with the optimal antenna spacing is 
\begin{equation}\label{eq:24}           C_{\textrm{dual}}=\sum_{i=1}^{2}\sum_{m=1}^{M}\log_2 \left({1+\frac{q_{ m,i} \mu_{i} \beta M }{\sigma^2}} \right),
\end{equation}
where the power allocation $q_{1,i},\ldots,q_{M,i}$ for $i=1,2$ is selected using the water-filling algorithm \cite{Telatar1999a}. Since there are two eigenvalues with multiplicity $M$, the power allocation is
\begin{align}
q_{m,1} &= \begin{cases} 
      \frac{P}{M}, &  P \le \frac{\sigma^2}{\mu_{2} \beta}-\frac{\sigma^2}{\mu_{1} \beta}, \\  
      \frac{P}{2M}+\frac{\sigma^2}{2\mu_{2} \beta M}-\frac{\sigma^2}{2\mu_{1} \beta M} & \textrm{otherwise},
   \end{cases} \\
q_{m,2} &= \begin{cases} 
      0, &  P \le \frac{\sigma^2}{\mu_{2} \beta}-\frac{\sigma^2}{\mu_{1} \beta}, \\  
      \frac{P}{2M}+\frac{\sigma^2}{2\mu_{1} \beta M}-\frac{\sigma^2}{2\mu_{2} \beta M} & \textrm{otherwise}. 
   \end{cases}
\end{align}
In the high-SNR regime where all power values are non-zero, the capacity expression in \eqref{eq:24} becomes
\begin{align} \nonumber
    C_{\textrm{dual}} &= M \log_2 \left(1+\frac{P \mu_{1} \beta  }{2\sigma^2}+\frac{ \mu_{1}-\mu_{2}}{2\mu_{2}}  \right) \\ 
    &+ M \log_2 \left(1+\frac{P \mu_{2} \beta  }{2\sigma^2}+\frac{ \mu_{2}-\mu_{1}}{2\mu_{1}}  \right). \label{eq:C-expression}
\end{align}
This capacity is exactly proportional to the number of dual-polarized antennas, $M$. This happens despite the fact that the total transmit power $P$ is divided over $2M$ eigendirections, because the receive beamforming gain also increases.

In the special case of perfect XPD (i.e., $\kappa=0$), we have $\mu_1 = \mu_2 = 1$ and the capacity expression in \eqref{eq:C-expression} simplifies to $2M \log_2 (1+\frac{P  \beta  }{2\sigma^2} )$. The multiplexing gain is then $2M$ and the transmit power is divided equally between the two polarizations. 
The channel matrix with the optimal spacing satisfies
$\tilde{\mathbf{H}}_{\rm d}^{\Htran}\tilde{\mathbf{H}}_{\rm d} = \beta M \mathbf{I}_{2M}$, which implies that the capacity is achieved by transmitting an independent signal from each antenna and from each polarization dimension.

The situation is different in the practical case of $\kappa > 0$.
When $\kappa$ increases towards $1/2$, the capacity is monotonically reduced because the imbalance between the eigenvalues in \eqref{eq:eigenvalues1}--\eqref{eq:eigenvalues2} increases. 
The Kronecker structure $\tilde{\mathbf{H}}_{\rm d}^{\Htran}\tilde{\mathbf{H}}_{\rm d} = \mathbf{K}^{\Htran}\mathbf{K} \otimes \beta M \mathbf{I}_M$ along with the eigenvalue decomposition in \eqref{eq:15} imply that independent signals should be transmitted from each of the $M$ antenna locations. However, the co-located dual-polarized elements are not transmitting independent signals. The strongest eigenvalue is achieved by transmitting the same signal from both polarizations, while the weaker eigenvalue is achieved by transmitting the same signal but with opposite signs using the two polarizations.
At low SNR, where all the transmit power is assigned to the eigenvalues $\mu_1 \beta M$, it is instead preferable to have a larger value of $\kappa$.
 
\subsection{Minimization of the array area or aperture length}
\label{subsec:minimize-area-aperture}

The capacity expressions in \eqref{eq:24} and \eqref{eq:C-expression} depend on the total number of antennas $M= M_{\rm h} M_{\rm v}$, but not on how these are divided between rows and columns in the URA. Hence, we can further optimize $M_{\rm h}$ and $M_{\rm v}$ while retaining the capacity.

The horizontal length $L_{\rm h}$ and vertical length $L_{\rm v}$ of the URA are calculated as
\begin{align}\label{eq:20}
    L_{\rm h}&=\Delta_{\rm h}(M_{\rm h}-1)+W = \sqrt{\frac{\lambda d}{M_{\rm h}}}(M_{\rm h}-1)+W, \\
    L_{\rm v}&=\Delta_{\rm v}(M_{\rm v}-1)+W = \sqrt{\frac{\lambda d}{M_{\rm v}}}(M_{\rm v}-1)+W,
\label{eq:21}
\end{align}
which is the distance between the outermost antenna locations plus the width $W>0$ of an individual antenna element.

Suppose we want to minimize the array area $L_{\rm h} L_{\rm v}$:
\begin{equation}\label{eq:22}
\begin{aligned}
& \underset{M_{\rm v},M_{\rm h} \in \{1,\ldots,M\}}{\text{minimize}}
& & L_{\rm h} L_{\rm v} \\
& \,\,\,\,\,\,\, \text{subject to}
& & M = M_{\rm h} M_{\rm v}.
\end{aligned}
\end{equation}
We can obtain an area expression that only depends on $M_{\rm h}$ by substituting $M_{\rm v}=M/M_{\rm h}$ into \eqref{eq:21}. The first derivative of $L_{\rm h} L_{\rm v}$ with respect to $M_{\rm h}$ then becomes
\begin{equation}
\frac{ W \sqrt{ \lambda d M_{\rm h}} \left(1 + M_{\rm h} -\sqrt{M} - \frac{M_{\rm h}}{\sqrt{M}} \right) + 2 \frac{\lambda d}{M} \left(M - M_{\rm h}^2 \right)}{2 M_{\rm h}^2}.
\end{equation}
The derivative equals zero when $M_{\rm h} = \sqrt{M}$, which corresponds to having a square-shaped array. However, this is a local maximum, while the area is minimized at either $M_{\rm h}=1$ or $M_{\rm h}=M$; that is, a ULA has the smallest possible area.

Suppose we alternatively want to minimize the total aperture length, namely the diagonal $\sqrt{L_{\rm h}^2+L_{\rm v}^2}$ of the URA. The corresponding optimization problem can be stated as 
\begin{equation}\label{eq:aperture-problem}
\begin{aligned}
& \underset{M_{\rm v},M_{\rm h} \in \{1,\ldots,M\}}{\text{minimize}}
& & L_{\rm h}^2+L_{\rm v}^2 \\
& \,\,\,\,\,\,\, \text{subject to}
& & M = M_{\rm h} M_{\rm v}.
\end{aligned}
\end{equation}
The minimum is then obtained when $M_{\rm h} = \sqrt{M}$, as can be shown by 
substituting $M_{\rm v}=M/M_{\rm h}$ into the cost function and equating its first derivative with respect to $M_{\rm h}$ to zero.

In summary, when deploying $M$ dual-polarized antennas to maximize the MIMO capacity, a one-dimensional ULA minimizes the aperture area while a square-shaped URA with $M_{\rm h}=M_{\rm v}=\sqrt{M}$ minimizes the aperture length.

\section{Numerical Results}

\begin{figure}[t!]
\includegraphics[scale=0.57]{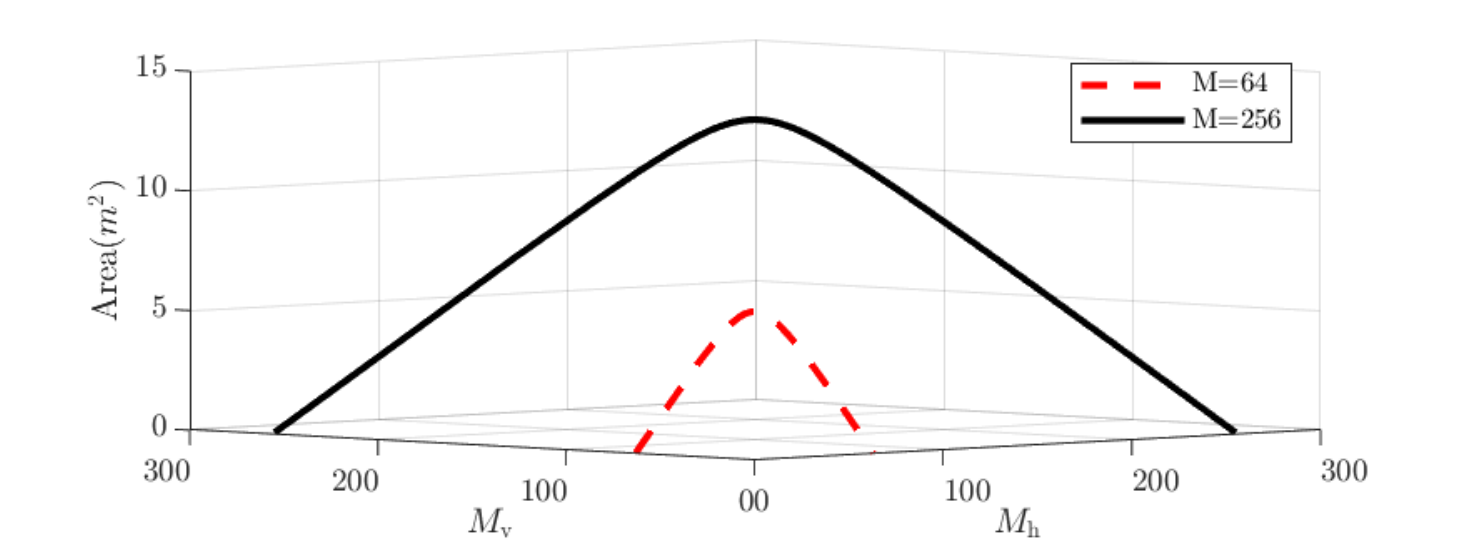}
\caption{The aperture area for varying $M_{\rm h}$ and $M_{\rm v}$ in the URA.}
\label{fig1}
\end{figure}

\begin{figure}[t!]
\includegraphics[scale=0.57]{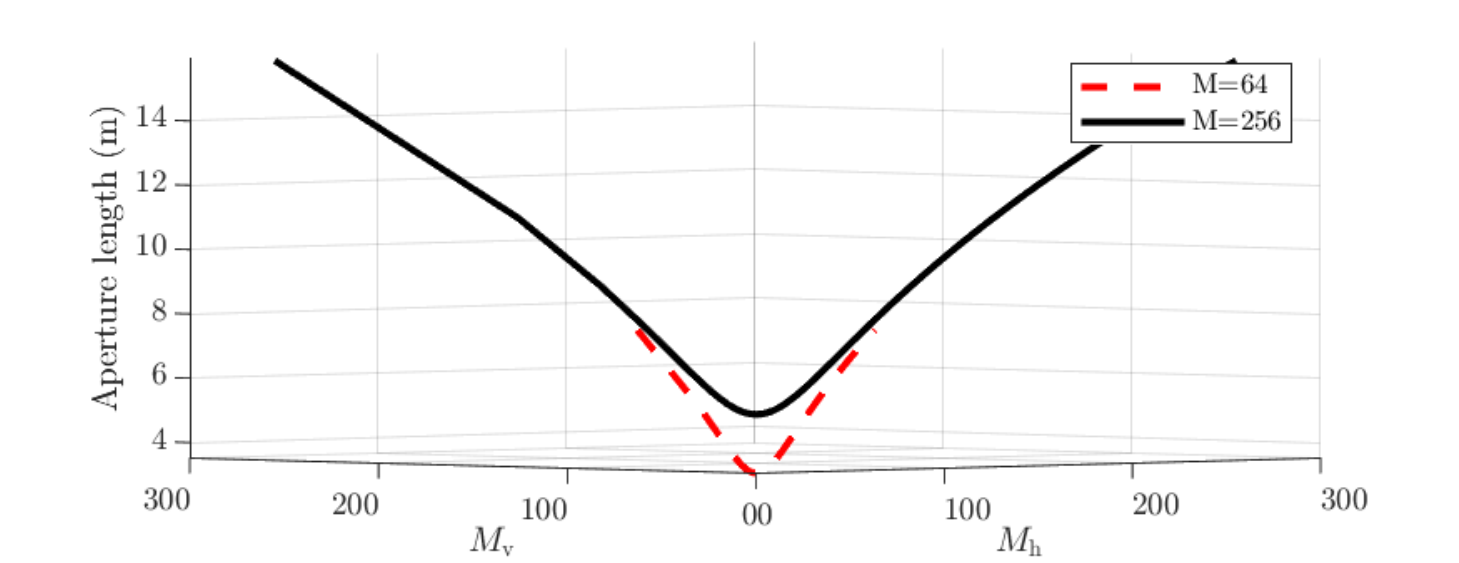}
\caption{The aperture length for varying $M_{\rm h}$ and $M_{\rm v}$ in the URA.}
\label{fig2}
\end{figure}

In this section, we provide simulation results that highlight the main results.
We consider a setup where the distance between the broadside transmitter and receiver antenna arrays is $d=100$ meters, with $W=\lambda/2$ meters. The carrier frequency is 30 GHz ($\lambda \approx 0.01$).
We begin by considering the physical dimensions of the URA when using the optimal antenna spacings. Fig.~\ref{fig1} shows the aperture area when varying $M_{\rm h}$ and $M_{\rm v}$, with either $M= M_{\rm h}M_{\rm v}=64$ or $M=256$.
The maximum area is obtained when $M_{\rm h}=M_{\rm v}$ (i.e., in the case of a USA), while the area decreases when the number of antenna elements increases in either the horizontal or vertical direction while decreasing in the other dimension. The smallest aperture area is obtained in the case of a horizontal/vertical ULA. 
The aperture length is shown in Fig.~\ref{fig2} for the same setup. We observe an inverse relation compared to the aperture area, so a USA obtains the minimum length with $M_{\rm h}=M_{\rm v}$. Both these results agree with the analysis in Sec.~\ref{subsec:minimize-area-aperture}.

\begin{figure}[htbp]
\includegraphics[scale=0.57]{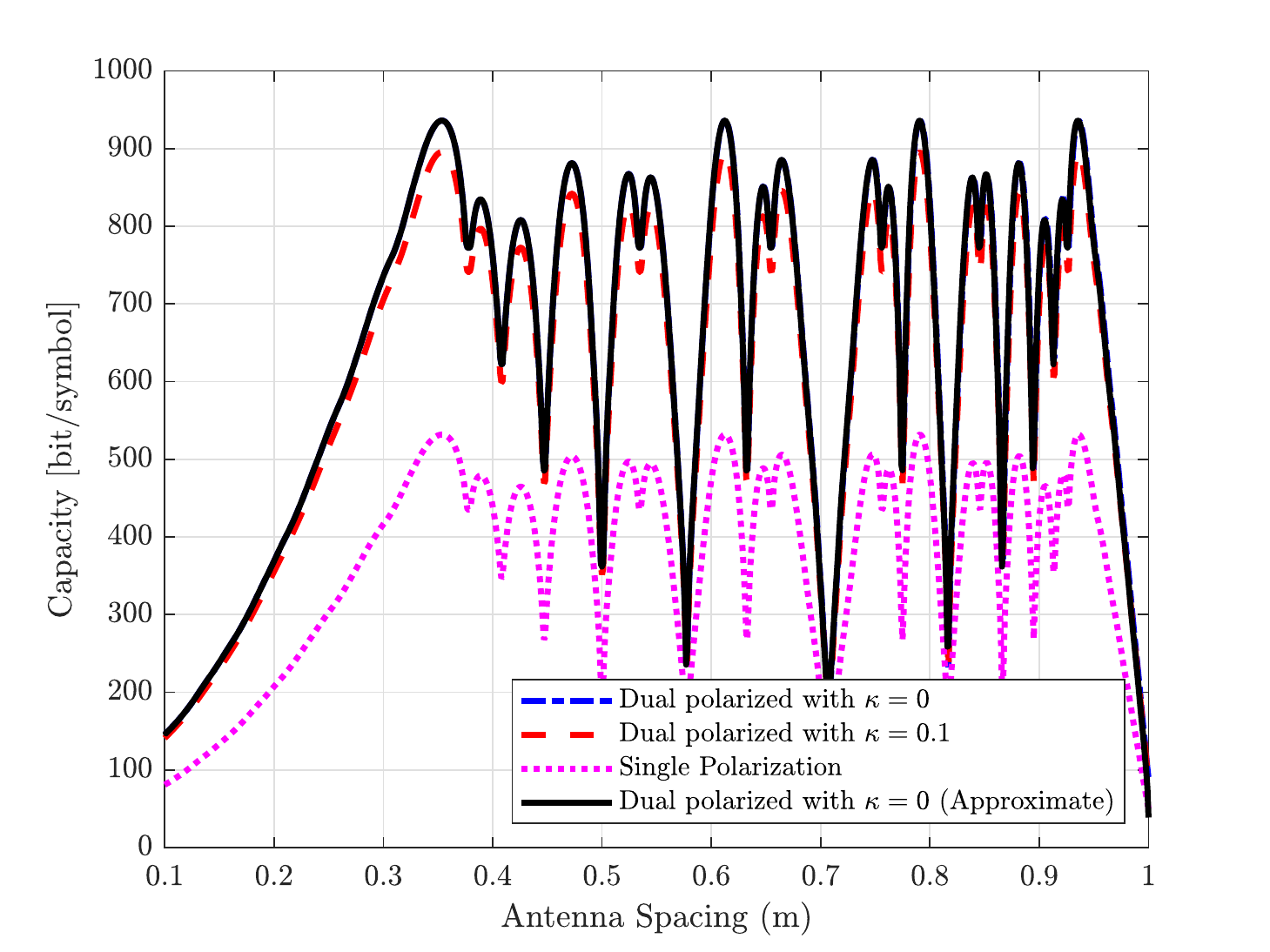}
\caption{The channel capacity as a function of the horizontal/vertical antenna spacing for a USA with $M_{\rm h}=M_{\rm v}=8$ antennas.}
\label{fig4}
\end{figure}

We will now demonstrate the relation between the antenna spacing and channel capacity in a MIMO setup with two USAs with $M=64$ dual-polarized antenna elements.  We consider an SNR of $P\beta/\sigma^2 =25$ dB in Fig.~\ref{fig4}. 
Increasing the antenna spacing first monotonically improves the capacity of the system, until we reach the maximum capacity at $\Delta_{\rm h} = \Delta_{\rm v} = 0.3535$\,m. At this point, the capacity crosses the threshold of $900$ bit/symbol, which is achieved using $128$ orthogonal spatial dimensions that each carries $7$ bit/symbol. When the antenna spacing is further increased, the capacity fluctuates and generally decreases. This highlights the importance of using the optimal spacing, while the price to pay is that each array is $2.5 \times 2.5$ meters at the optimal point in this setup. The XPD leads to a minor capacity reduction, as can be seen by comparing the curves with $\kappa=0$ with $\kappa=0.1$. Most results in this figure are obtained using the exact channel matrix model in \eqref{eq:H_d_def}, but the solid line is obtained using the approximation in \eqref{eq:Hu_approx} that was used to obtain analytical results. The tightness of the approximation is clearly visible, and the approximate curve coincides with the actual curve. Finally, we notice that the capacity is nearly doubled when using dual polarization compared to using a single polarization.

\section{Conclusion}
In this paper, the MIMO channel capacity was maximized for a given propagation distance between two dual-polarized URAs. The optimal spacing between the antenna elements was identified analytically. The trade-off between the aperture length and the area enclosed by the array was manifested analytically and illustrated via simulations. A ULA requires the smallest area for a given number of antennas, while the USA gives the smallest aperture length. The cross-polar discrimination determines how to transmit optimally using dual polarization, but even when it is imperfect, the capacity is nearly doubled compared to using a single-polarized array.   

\bibliographystyle{IEEEtran}
\bibliography{IEEEabrv.bib,bibliograph.bib}
\end{document}